\documentclass[runningheads]{llncs}
\usepackage{graphicx}

\usepackage{tikz}
\usepackage{comment} 
\usepackage{amsmath,amssymb} 
\usepackage{color}

\usepackage{comment}
\usepackage{booktabs}
\usepackage{xspace}
\makeatletter
\DeclareRobustCommand\onedot{\futurelet\@let@token\@onedot}
\def\@onedot{\ifx\@let@token.\else.\null\fi\xspace}

\def\ie{\emph{i.e}\onedot}

\makeatother

\begin{document}
\pagestyle{headings}
\mainmatter

\title{Toward Smart Doors: A Position Paper} 


\titlerunning{Toward Smart Doors: A Position Paper}
%
\author{
Luigi Capogrosso    \and
Geri Skenderi       \and
Federico Girella    \and
Franco Fummi        \and
Marco Cristani
}
\authorrunning{L. Capogrosso et al.}
%
\institute{Department of Computer Science, University of Verona \\
\email{\{name.surname\}@univr.it} \\}
\maketitle

\begin{abstract}
Conventional automatic doors cannot distinguish between people wishing to pass through the door and people passing by the door, so they often open unnecessarily. This leads to the need to adopt new systems in both commercial and non-commercial environments: \emph{smart doors}. In particular, a smart door system predicts the intention of people near the door based on the social context of the surrounding environment and then makes rational decisions about whether or not to open the door.
This work proposes the first position paper related to smart doors, without bells and whistles. We first point out that the problem not only concerns reliability, climate control, safety, and mode of operation. Indeed, a system to predict the intention of people near the door also involves a deeper understanding of the social context of the scene through a complex combined analysis of proxemics and scene reasoning.
Furthermore, we conduct an exhaustive literature review about automatic doors, providing a novel system formulation. Also, we present an analysis of the possible future application of smart doors, a description of the ethical shortcomings, and legislative issues.
\end{abstract}

\section{Introduction}
\label{sec:intro}


An \emph{automatic door}, also known as an auto-door, is a door that opens automatically by sensing the entrance of a person or an object in a small area close to the door. Automatic doors have been in existence for many centuries. We have to go back to the first century AD in Roman Egypt. It was here that a mathematician, Heron of Alexandria, invented the first known automatic door: a mechanism employed to open a temple’s doors when a fire was lit at the altar~\cite{papadopoulos2007heron}.

In the last decade, automatic doors are found in many different places, spanning from airports, residential areas, banks and hospitals, to offices and industrial units. They automatically detect the presence of an agent (a person or a big object) nearby, using sensors (infrared, ultrasonic) and other components. Even though these techniques are effective and successful in detecting people, they fail to understand their intention to enter or exit with respect to the surrounding environment~\cite{lwin2015automatic}.

In these scenarios, we can identify two main challenges. The first one is that an automatic door cannot decide whether it should open or not based on the people's intention, especially when the person is approaching quickly~\cite{sumbul2011control} (if a person just passes by a door and has no intention of entering it, the door will open unnecessarily). The second one is that an automatic door cannot distinguish between the different actors that can interact with it ~\cite{yang2013intelligent} (we might not want to open the door for a 2-year-old child).

\begin{figure}[t]
\begin{center}
\includegraphics[width=0.9\linewidth]{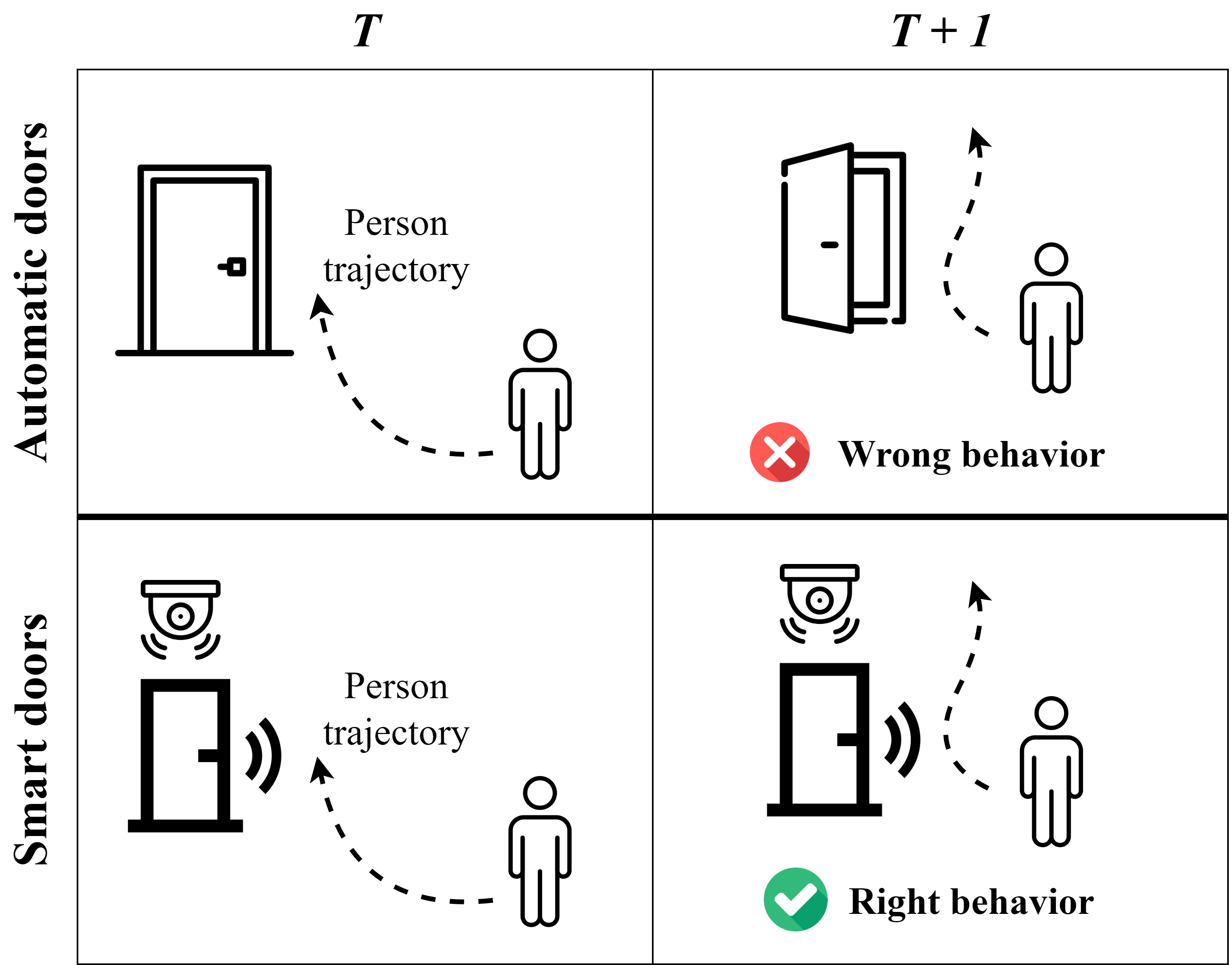}
\end{center}
\caption{\emph{Smart door} teaser: the image shows the behavior of a conventional and a smart door in the case of a tangent trajectory by a person. In none of the depicted examples there is an intention to enter the door. As we can see, the behavior of conventional automatic doors at instant $T+1$ is wrong, \ie{}, the door opens at the moment the person starts to approach, creating a false positive. On the other hand, the smart door, predicts the intention of the people approaching the door, and by rationally deciding whether or not to open the door, understands that the person does not want to enter (only passing nearby) and does not open.}
\label{fig:Figure1}
\end{figure}

Based on what has been said so far, there is a clear need to incorporate the decision-making process to make automatic doors intelligent: we are moving into the era of \emph{smart doors}. Let's define a smart door as a system that predicts the \emph{intention} of people near the door based on the social context of the surrounding environment and then makes rational decisions about whether or not to open the door. Eventually, it should also hold the door open for several seconds until no person is in the detection area, and then close.

To ease the transition to smart doors, researchers need to provide these systems with a sort of ``human intelligence''. Therefore, the key is to add machine/deep learning algorithms to automatic doors. In particular, automatically predicting future movements of people is a hot topic in the pattern recognition field~\cite{lin2021survey}. Applying this, in a nutshell, gives the door system the ability to understand humans by following a pedagogical~\cite{kiru2020intelligent} and psychological~\cite{walther2006interactions} process. In fact, a smart door system able to predict the intention of people near the door requires a complex combination of \emph{proxemics} and \emph{scene reasoning}: the former considers that individuals behave differently if they are alone, in a group, or a crowd. The latter takes into account the constraints of the scene (people cannot cross the walls).

Furthermore, the use of smart doors with synergies among other automation systems (motorized lock and pressure pad) can allow for the remote management of door access, climate control, and safety. It could be argued that the realization of these concepts will be most significant toward promoting independent living and increasing the quality of life for the rising number of older and impaired people, with the prevalence of chronic conditions such as cardiovascular disease and dementia~\cite{donnelly2012impact}.

Smart doors can also be used as a cost-effective and reliable method of using machine/deep learning algorithms and sensors to build a healthy environment~\cite{varshini2021iot} (a smart entry device in a shop center that automatically counts people, monitors human body temperature, and detects a mask at the door opening system).

Finally, we emphasize that smart doors are ideal for commercial and public sector buildings and installations that may involve a large number or variety of users, such as able-bodied, disabled, young, and elderly~\cite{lymperopoulos2014pathpass}. There is an absolute need to ensure safe and easy access for all, yet special attention must be paid to the application of these systems~\cite{robles2010review}. This requirement impacts architects and specifiers who, in addition to meeting the conflicting needs of a range of stakeholders, must also ensure that legal duties are met to minimize the risks of dangerous situations.
\section{Related Work}
\label{sec:related}

The automation of doors has seen numerous approaches and evolution throughout the years, with smart doors being the latest iteration in the innovation process. This section focuses on the research efforts that precede this work. It includes studies dedicated to several automatic door systems proposed by researchers in previous years and the first attempts that use machine/deep learning in developing smart door systems. These deep learning models are trained using datasets like the ones we will introduce in Sec.~\ref{subsec:rel_tad}, allowing smart doors to learn automation through proxemics and scene reasoning.

The different automation approaches can be grouped into the following four different categories: \emph{(i)} automation through sensors, \emph{(ii)} automation through detection, \emph{(iii)} automation through intention, and \emph{(iv)} automation through proxemics. For each one, we will indicate their methodologies, strengths, and weaknesses.

\subsection{Automation Through Sensors} 
Automation through sensors is the most common approach. It uses sensors, such as motion or infrared, to trigger the door opening. This category has seen some evolution over the years, particularly in the use of different sensors: starting with classic infrared (IR) sensors~\cite{garcia2018LowPowerSensor}, the methodology has progressed by using other sensors to improve performance. For example, in~\cite{song2009development} and~\cite{kim2017object} the use of ultrasonic sensors brought an improvement in motion detection, especially in outdoor scenarios where IR light from the sun would interfere with the door's IR sensor.

More recently, there has been a shift to the use of temperature sensors: these open the door if and only if the temperature does not exceed a maximum threshold~\cite{iskandar2021automatic}. This method was invented and used to reduce the spread of COVID-19 virus by avoiding opening the door to feverish people~\cite{supriana2021implementation}.

Unfortunately, this category of approaches produces a high number of false positives (unnecessary door openings) since they do not classify the subjects generating the detected motion. This causes the door to open even when it is not necessary if the sensors detect motion.

\subsection{Automation Through Detection}
With detection, we refer to the task of recognizing people in a custom scenario (see Sec.~\ref{subsec:cv} for more details). With this definition, we can describe the approach in this category as using cameras such as RGB or RGB-D to detect people and open the door.

The most common approach is based on cameras whose output is sent to a machine learning detection model, whose task is to predict whether the scene contains a person. For example, such an approach is presented in~\cite{fosstveit2012intelligent} and~\cite{feng2017control}, where a Kinect and its API are used to detect people in the scene and open a door accordingly. Another similar implementation is presented in~\cite{lwin2015automatic}, where a camera is used to perform face detection and recognition.

While reducing the number of false positives compared to the sensor-based method, this category still suffers from false positives, for example, when a person walks past the door without intending to walk through it. Therefore, the intention is a crucial factor to consider when the goal of the system is oriented toward reducing the number of false positives.

\subsection{Automation Through Intention}
As mentioned several times above, a person may walk past a door without intending to go through it. In such cases, an automatic door should not open, and if it does, it should be considered a false positive, meaning that the system incorrectly predicted that the person would need to open the door. The use of intention-based systems has seen a recent rise in popularity, mainly due to the increase in computing power that allows for the training of a model and the use of more complex algorithms in a real-time scenario.

Although studying people's intentions is difficult, it can be made easier by making them explicit to users. In~\cite{jimoh2007intelligent} and~\cite{wu2016design} a speech recognition model is used to recognize user commands and open doors accordingly. Another approach is to make intentions explicit by asking the user to show a token~\cite{yoon2020iot}, assume a predefined pose~\cite{terashima2021proposal}, or perform a predefined gesture~\cite{chiu2021gesture}, all of which must be recognized by machine learning models.

While these approaches reduce the number of false positives compared to previous ones, they introduce the risk of increasing the number of false negatives when the door does not open due to misidentification of the necessary commands caused by the innate complexity of these tasks. Another problem with these systems is the need for explicit user interactions, which makes them unusable in an everyday life scenario where users may not be able to perform the necessary gestures or people would not be able to know what gesture to make.

\subsection{Automation Through Proxemics and Scene Reasoning}
To ease the transition to smart doors and predict the intention of people approaching the door, without them having to perform tasks out of the ordinary, a system capable of performing a combination of proxemics and scene reasoning is needed.

One of the first attempts in this direction is the one made by us in~\cite{skenderi2021dohmo}. The system aims at assisting elderly and impaired people in co-housing scenarios, in accordance with privacy design principles. This approach estimates user intention through a trajectory forecasting algorithm, predicting the trajectory of people and using this information to infer the intention to use (or not use) the door. These methods result in a seamless experience for the user, who no longer has to worry about interrupting their workflow to issue commands to the system. However, some aspects of this work need to be improved. The first is that the proposed system focuses only on the analysis of elderly and disabled people. The second is that to use these systems we need datasets on which to train our models, and state-of-the-art research has produced only a few benchmarks to date, as we will see in the next section.

\subsection{Testing Automatic Doors}
\label{subsec:rel_tad}
All systems that implement the methods presented above must undergo a validation process to confirm that they work properly. For the traditional automatic doors (with sensors), the validation process generally evaluates the performance of the sensors, motors, and safety measures for a given door. These performances need to comply with a set of guidelines to ensure the safety of operations (as discussed in Sec.~\ref{sec:legislation}).

However, for smart doors, this validation process involves additional steps given the introduction of a machine or deep learning models. These models, as presented in the previous section, are used to assess the presence and intention of people and operate the doors accordingly. In order to train these models and validate their performance offline, we require datasets.

For smart doors models, datasets are composed of videos and images accompanied by ground truth annotations, meaning that each element of the dataset is annotated with the correct behavior expected from the model. For example, a model trained to detect people through a camera will use a dataset containing images (or videos) of people in a similar situation to the one it will be installed in. 

Especially for deep learning models, large datasets are needed to perform meaningful training. As such, deep learning models have seen limited application to trajectory forecasting. In~\cite{styles2019forecasting}, the authors address the lack of training data by introducing a scalable machine annotation scheme that enables our model to be trained using a large dataset without human annotation.

In~\cite{fatima2013analysis}, a framework is developed for smart homes dataset analysis to reflect their diverse dimensions in a predefined format. It analyzes a list of data dimensions that covers the variations in time, activities, sensors, and inhabitants.

As we can see, the state of the art regarding datasets to train intelligent models for smart door systems is also still in a flux. We note a surprising absence of indoor trajectories datasets. The only work that comes close in this regard is~\cite{fatima2013analysis}, whose results are useful for upcoming researchers to develop a better understanding of the characteristics of smart home datasets and classifier performance.

\section{Toward Smart Doors}
\label{sec:toward_sm}

The creation of a smart door requires solving classical computer vision and forecasting tasks, namely people detection/tracking (Sec.~\ref{subsec:cv}) and trajectory forecasting (Sec.~\ref{subsec:forecasting}).

In the following, we will describe these two modules in detail re-targeting, when possible, previous computer vision and forecasting methods that can provide a solution to these problems.

\subsection{People Detection and Tracking}
\label{subsec:cv}
Detecting and tracking people is an important and fundamental component of many interactive and intelligent systems~\cite{liciotti2017people}. In particular, the problem of detecting a person can be stated as: given a video sequence or an image, localize all the subjects that are humans~\cite{brunetti2018computer}. So then, for tracking people, we start with all possible detections in a frame and assign them an ID, which we try to carry forward into subsequent frames. If the person has moved out of the frame the ID is dropped, and if a new person appears, we assign them a new identifier~\cite{yilmaz2006object}.

Both detection and tracking are challenging problems, especially in complex real-world scenes that commonly involve multiple people, changes of appearance, complex and dynamic backgrounds, and complicated occlusions~\cite{andriluka2008people}. Nonetheless, people detection has reached impressive performance in the last decade given the interest in the automotive industry and other application fields~\cite{benenson2014ten}. At the same time, people tracking systems have become increasingly robust even in in crowded scenes due to the extensive efforts in video surveillance applications~\cite{fuentes2006people}.

Different people detection techniques have been designed to work indoor~\cite{tseng2014real}, outdoor~\cite{jafari2014real} and in crowded scenes~\cite{wang2018repulsion}. When the image resolution becomes too low to spot single people, regression-based approaches are used~\cite{boominathan2016crowdnet,chan2008privacy}, also providing, in some cases, density measures~\cite{liu2018decidenet,xu2016crowd}.

In general, the process of detecting people from video (or images) can be performed with the following four sequential steps: \emph{(i)} extracting candidate regions that are potentially covered by human objects~\cite{dalal2006finding}, \emph{(ii)} describing the extracted region~\cite{nguyen2016human}, \emph{(iii)} classifying or verifying the regions as human or non-human~\cite{zhao2000stereo}, and \emph{(iv)} post-processing (merging positive regions~\cite{dalal2005histograms} or adjust the size of those regions~\cite{leibe2005pedestrian}).

Once the people within the image have been identified, if the work imposes the assumption that the ground plane is planar, it is necessary to estimate a homography given by some reference features or using information from the vanishing points detected in the image~\cite{criminisi2000single,rother2002new}.

On the other hand, a simple method for tracking multiple objects is using the Euclidean distance between the centers of bounding boxes at successive frames. Thus, an object in a frame is considered the same as an object in a previous frame if the Euclidean distance between their bounding box centers is small~\cite{cojocea2020efficient}. An improvement to this method can be made using a Kalman filter~\cite{welch1995introduction}, which is able to predict the future position of objects based on their past trajectory. These presented are very simple tracking methods, which therefore do not allow for the accurate implementation of a smart door system. So, it is necessary to use more advanced systems such as DeepSORT~\cite{wojke2017simple} that do not rely solely on the objects position and movement but introduce a metric which compares two people regarding their visual features.

\subsection{Trajectory Forecasting}
\label{subsec:forecasting}
Forecasting people's trajectories has been a topic of studies for over two decades. For an in-depth dive into the detailed methodological aspects, the interested readers can refer to various surveys~\cite{becker2018evaluation}, \cite{morris2008survey}, and~\cite{franco2022under}. In its essence, trajectory forecasting is a multivariate time-series forecasting task, considering that in most applications the system has to predict the $x$ and $y$ coordinates of a tracked object of interest over time. This means it is necessary to jointly forecast the $x$ and $y$ signals and understand the intricate relationship between them. These predictions can be made in a probabilistic manner or by simply estimating the expected values (point forecasting). In the first case, we try to forecast the parameters of the assumed distribution of the future trajectory. Because this is often assumed to be a Gaussian, the predictions are therefore the respective means and the diagonal covariance matrix. 

In the literature, there are two main aspects that go into modelling trajectories in an optimal manner: the first one is using or producing well-performing sequence models and the other is using multi-modality and modeling the social information available in the scene. The impressive performance of gated recurrent neural networks such as LSTMs~\cite{hochreiter1997long} and GRUs~\cite{cho2014learning} on sequential tasks quickly made its way into trajectory forecasting, with several pioneering works~\cite{alahi2016social,salzmann2020trajectron++} using these RNN variants. Little after their massive success in NLP, the Transformer~\cite{vaswani2017attention} made its way into trajectory forecasting as well~\cite{giuliari2021transformer}. To take another step further in modeling social interactions, the seminal work of~\cite{gupta2018social} proposed to use a GAN-based approach and generate plausible trajectories.

Large improvements in trajectory forecasting came with endpoint-conditioned models, which try to estimate the final goal of the moving agent and then its trajectory~\cite{mangalam2020goals,mangalam2021goals}. We argue that in the case of smart doors, this approach is crucial to allow the application of intelligent doors. The reasoning behind this is that the social interactions in a closed world are much more limited than in open scenarios, which are the cases considered in the benchmark dataset of all the works cited above. There is of course less variability in an office than out on the streets, but much less space. To provide an optimal trajectory estimation, the model needs to estimate where the agent is heading first. This is even more important for smart doors since the goal estimation is already half of the job: If the system knows the agent will reach the door, then it is simply a matter of understanding how many time steps it will take \ie{}, knowing its trajectory. Probabilistic forecasting of trajectories is important for indoor scenarios and smart doors, as it is for its outdoor counterpart. In industrial settings, most decisions will be made based on the confidence interval of the trajectory. We believe that the advantage of being indoors is that, given an optimal forecasting model, there is little variance to particular trajectory types and therefore the corresponding confidence intervals will be relatively narrow. This is turn translates to high-confidence decisions.

The problem can also be tackled in alternative ways, as demonstrated by~\cite{skenderi2021dohmo}, where we try to cast the trajectory forecasting problem directly as a time-series classification task. This is done by labeling different trajectories that are extracted via a detection and tracking module from a camera inside a room. This approach produces excellent results utilising a very efficient Random Forest model, which also provides interpretability for shorter trajectories. In general, however, the acquisition of trajectories must take place at an adequate distance to allow the system to make the prediction and consequently then open or not open the door. By considering the problem purely as a trajectory forecasting task, the decision then becomes a variable that can be decided based on the industrial set-up, which provides even more flexibility. These learning-based approaches open up a world of new possibilities for smart doors in large commercial settings, creating energy-efficient solutions.

\section{Legislative Issues}
\label{sec:legislation}

The implementation and use of automatic doors are subject to legislation, enforcing guidelines on the installation and correct behavior needed to ensure easy and safe interactions with users under different circumstances. We argue that smart doors, seen as an evolution of the conventional automatic door, must also comply with these guidelines and that the presence of this legislation demonstrates the feasibility of the proposal of this position paper. Additionally, we will address the privacy aspect of these smart systems in Sec.~\ref{sec:discussion}, introducing ways to avoid privacy concerns over the data being used. Given that laws may vary from country to country, we will only mention general aspects that can be taken into account in the software used to operate the doors, as the production and physical installation of the doors are beyond the scope of our work.

The most important aspect is the \emph{overall safety} of the users during daily operations. It includes safety for the users operating the door as well as people in the vicinity of the door, defining areas of operation in which the system needs to take care to not cause harm.
The first area of operation is the one users cross to reach other rooms. A system needs to be able to detect people in this area to avoid closing the door and injuring them. A combination of pressure and IR sensors is usually present in automatic door systems, allowing the system to detect people and revert to the open state.
The second one is the area of operation of the door itself. In sliding doors, for example, this includes the area in which the door slides while opening. In a revolving door, this is the area where the door swings while opening. If people find themselves in these areas they risk being hit, crushed, or severed by the door itself. Many laws are in place enforcing physical boundaries for the regions surrounding these areas. In European legislation, laws are in place to limit the maximum speed of automatic doors, limiting the damage to users being hit. Other laws define distances between the doors and other hard surfaces, limiting the risk of crushing.

Finally, another crucial aspect is \emph{behavior during emergencies}. With emergencies, we refer to situations in which either an emergency signal is being sent to the door (manually or by another system), or the system detects that it can no longer operate correctly. An example of this may be a fire alert, where an outside system (or an operator) signals the door that an emergency is taking place. Another instance of an emergency might be a power outage, where no signal is being sent but the door can no longer operate normally. In these cases, according to European legislation, doors must switch to a safe state that allows them to be used even in the absence of power. This safe state can be an open state, in which the door remains open, or a manual state, in which the door can be operated manually.

Another similar issue is related to the \emph{detection of malfunctions} in the system (not opening the door when needed). In this case, the system needs to detect its malfunction and change its behavior accordingly, for example by default to the safe states previously mentioned.

All these problems can be further mitigated through proper implementation of the smart door control system. Interaction with emergency systems is clear, and default safe states can be implemented even without power. Proper detection of people in the areas of operation (using the method presented in Sec.~\ref{subsec:cv}) can prevent smart doors from opening and closing when they detect people in these areas, aiding traditional sensors and further limiting the risk of injuries. As such, we argue that smart doors, as presented in this work, would not only follow the requested legislation but provide additional safety features, thanks to their control system and use of cameras.

\section{Discussion}
\label{sec:discussion}

Major advances in computer vision have set the stage for the widespread deployment of connected devices with \emph{always-on} cameras and powerful visualization capabilities. While these advances have the potential to enable a wide range of novel applications and interfaces, such as smart doors, privacy and security concerns surrounding the use of these technologies threaten to limit the range of places and devices where they are adopted. 

Optical cameras can be the most suitable sensors for a smart door system, but the acceptance of this technology can be difficult since it raises privacy concerns. Video footage can reveal the identity of the people being filmed, and in general, nowadays, recording is regulated by strict laws, both nationally and internationally. In addition, potential attacks on video transmission channels and storage servers can be a serious security problem.

We believe that, despite the widespread cloud-centered application, this should not be the only paradigm to refer to and that a new approach to decentralization is needed. In particular, by adopting the \emph{edge-centric computing} paradigm, we will push the frontier of computing applications, data, and services away from centralized nodes to the periphery of the network. This allows the implementation of the \emph{privacy-by-design} principle, as is shown in~\cite{skenderi2021dohmo}. So, first, processing the video internally (through an edge device) and then, transmitting only the result of the computation to the remote server, without any sensible information.

While great strides have been made in lowering the power requirements of on-board processing by novel, highly efficient microcontrollers~\cite{gautschi2017near}, one of the most severe energy efficiency bottlenecks is the need to periodically process data found in traditional polling-based systems~\cite{scherer2021towards}. One approach to reducing the amount of data that needs to be processed is \emph{event-driven processing}, where data is only processed if certain activity conditions are met, like motion in a video stream.

At the same time is worth nothing that current computer vision technology is now mature enough to handle privacy issues. In addition, the legislative and legal conditions are also extremely specific in order to minimize any risks to people as much as possible. So, through the intersection of different research fields, combining their respective paradigms and technologies, nowadays it is possible to create a smart camera system that is fully secure from the perspective of privacy and ethics.

\section{Conclusion}
\label{sec:conclusion}

In this position paper, we introduce smart doors as systems that predict the intention of people near the door, following the social context of the surrounding environment, and making rational decisions relative to whether or not to open. We have seen how social context requires such systems to optimize processes like climate control and security or develop emerging application scenarios to improve our quality of life.

In addition, the implementation of these systems allows for the analysis of social context, and thus advances research in this sociological regard as well. Smart doors cover to date, and will do so even more in the future, an important role in several fields, thus providing a continuous source of interest in this multidisciplinary problem.

\paragraph{\textbf{Acknowledgements.}} This work has been partially supported by the Italian Ministry of Education, University and Research (MIUR) with the grant ``Dipartimenti di Eccellenza'' 2018-2022.

\clearpage
%
%
\bibliographystyle{splncs04}
\bibliography{egbib}
\end{document}